\pdfoutput=1

\documentclass[conference]{IEEEtran}
\usepackage[utf8]{inputenc}
\usepackage[T1]{fontenc}
\IEEEoverridecommandlockouts

\usepackage{cite}
\usepackage[a4paper, top={0.68in}, bottom={0.68in}, left={0.68in}, right={0.68in}]{geometry}
\usepackage{textcomp}
\usepackage{xcolor}
\usepackage{times,url}
\usepackage{epsfig}
\usepackage{algorithm}
\usepackage{algorithmic}
\usepackage{graphicx}
\usepackage{subcaption}
\usepackage{amsmath}
\usepackage{amssymb}

\usepackage{pgfplots}
\pgfplotstableset{col sep=comma}

\pgfplotsset{
  poweraxisRaw/.style={
    major grid style={line width=.2pt,draw=gray!50},
    grid=both,
    width=8cm,
    height=4cm,
    legend style={nodes={scale=0.7, transform shape}},
    ymin=0.0, 
    ylabel={Energy (kWh)},
    xlabel= {Time (days)},
    yticklabel style={align=right},
    xtick={0,48,96,144,192,240,288,336,384,432,480,528,576,624},
    xticklabels={M,T,W,T,F,S,S,M,T,W,T,F,S,S},
  },
  poweraxis/.style={
    poweraxisRaw,
    ymax= 1.5,
    xlabel= {Time (days)},
    x label style ={yshift=-\baselineskip, anchor=south},
    y label style = {yshift=-0.5cm, anchor=south},
    xmin=0,xmax=672,
  },
  poweraxisHalf/.style={
    poweraxis,
    xmin=0,xmax=336,
    xtick={0,48,96,144,192,240,288},
    xticklabels={M,T,W,T,F,S,S},
  },
  poweraxisRevision/.style={
    poweraxisRaw,
    width=8cm,
    height=4cm,
  },
  powerplot/.style={
    mark=none,
  },
  table/powertable/.style={
    x = b,
    y = a,
    col sep=comma,
  },
  table/powertable0/.style={
  powertable,
  y = Real
  },
  table/powertable1/.style={
  powertable,
  y = Predicted
  },
}

\newcommand{\findmax}[3]{
  \pgfmathsetmacro{\mymax}{0.0}%
  \pgfplotstableforeachcolumnelement{#2}\of#1\as\cell{%
    \ifthenelse{\lengthtest{\cell pt > \mymax pt}}%
    {\pgfmathsetmacro{\mymax}{\cell}} %
    {}%
  }%
  \pgfmathsetmacro{#3}{\mymax}%
}

\newcommand{\indicators}{indicators}
\newcommand{\avgII}{Average Indicator Distance}

\newcommand{\powerplotComparison}[1]{%
  \findmax{plots/Sample#1.csv}{a}{\maxa}%
  \findmax{plots/Example#1.csv}{a}{\maxb}%
  \pgfmathsetmacro{\ymax}{1.1 * max(\maxa,\maxb)}%

  \begin{tikzpicture}
    \begin{axis}[poweraxis,
      ymax = \ymax,
      ylabel = {Generated},
      xlabel = {},
      name=first,
      ]
      \addplot +[powerplot,] table [powertable]{plots/Sample#1.csv};
    \end{axis}
    \begin{axis}[poweraxis,
      at=(first.below south west),
      yshift=-.8\baselineskip,
      ymax = \ymax,
      ylabel = {Natural},
      anchor=north west,
      ]
      \addplot +[powerplot] table [powertable]{plots/Example#1.csv};
    \end{axis}
  \end{tikzpicture}
}

\newcommand{\oldversion}[1]{}

\newcommand{\newversion}[1]{#1}

\begin{document}

\title{Privacy-Preserving Synthetic Smart Meters Data}

\author{\IEEEauthorblockN{Ganesh Del Grosso}
\IEEEauthorblockA{\textit{INRIA, Ecole Polytechnique} \\
Gif-sur-Yvette, France \\
ganesh.del-grosso-guzman@inria.fr}
\and
\IEEEauthorblockN{Georg Pichler}
\IEEEauthorblockA{\textit{Technische Universität Wien} \\
Vienna, Austria \\
georg.pichler@ieee.org}
\and
\IEEEauthorblockN{Pablo Piantanida}
\IEEEauthorblockA{\textit{CentraleSup\'elec, CNRS, Universite Paris-Saclay}\\
Gif-sur-Yvette, France \\
pablo.piantanida@centralesupelec.fr}

}
\maketitle

\begin{abstract}
Power consumption data is very useful as it allows to optimize power grids, detect anomalies and prevent failures, on top of being useful for diverse research purposes. However, the use of power consumption data raises significant privacy concerns, as this data usually belongs to clients of a power company. As a solution, we propose a method to generate synthetic power consumption samples that faithfully imitate the originals, but are detached from the clients and their identities. Our method is based on Generative Adversarial Networks (GANs). Our contribution is twofold. First, we focus on the quality of the generated data, which is not a trivial task as no standard evaluation methods are available. Then, we study the privacy guarantees provided to members of the training set of our neural network. As a minimum requirement for privacy, we demand our neural network to be robust to membership inference attacks, as these provide a gateway for further attacks in addition to presenting a privacy threat on their own. We find that there is a compromise to be made between the privacy and the performance provided by the algorithm.
\end{abstract}


\section{Introduction}

\newversion{Power consumption data from individual households is essential for diverse research and operational purposes. It is already abundant and with the help of cheap smart metering devices very fine-grained analysis becomes possible.
  As machine learning (ML) algorithms can be implemented with ease, even on consumer hardware, and ML competitions flourish, energy providers might be tempted to take advantage of this opportunity. If an anonymized portion of their collected power consumption data is released publicly, scientists around the world can utilize it and they might uncover patterns or mechanisms that aid system operations.
  However, energy consumption data cannot be published or released, even when anonymized, without potentially violating the privacy of individuals. An attacker with enough side information to identify a particular household might be able to discern information, such as geographic location, active hours and even the specific appliances \cite{fernandez2016online} present.

  Our aim is to develop the means for generating synthetic power consumption data that faithfully imitates the important statistical properties of original data as an ensemble, even including variations and outliers. Then, this synthetic data can be used instead of real data for the planning of system operations and, as it is not tied to specific individuals, it can be distributed without privacy concerns.
  To approach this problem, we utilize Generative Adversarial Networks (GANs)~\cite{goodfellow_generative_2014}, where, based on real power consumption data, an artificial neural network is trained to generate synthetic curves. This makes our task twofold: On the one hand, we need to carefully evaluate the generated data for its utility and on the other hand, we also need to assure that the generated curves do not enable an attacker to draw conclusions about the individuals present in the original training set.

To address the issue of performance evaluation, two different methods are proposed. The first evaluation method relies on comparing the relevant statistical properties of artificially generated data to those of natural data. The second revolves around using artificial data for a prediction task

and comparing the results to those obtained by using natural data instead.

A Long Short-Term Memory (LSTM) network is trained for this purpose. This way of addressing the quality of generated data is common practice with GANs in the context of images, where artificial images are used to train a classifier network and the performance of this classifier is compared to a baseline trained with natural images \cite{borji_pros_2018}.

Regarding privacy, membership inference attacks directly measure the privacy leakage of a machine learning model \cite{nasr_comprehensive_2018}. In view of this, we utilize three mechanisms to perform membership inference attacks, that are model agnostic within the GAN framework.

Different setups are considered, where we study the trade-off between privacy and performance of our algorithm by tuning the training process of the GAN.
}

\oldversion{Power consumption data from individual households is  valuable information for planning and operating a power grid. However, these readings constitute personal data which  could potentially reveal sensitive information when released. An attacker with access to the power consumption curves of a household could determine sensitive information, such as the active hours and appliances present. To address this problem, we propose the alternative of using synthetic power consumption data that faithfully imitates the important statistical properties of real data. This artificial data can be used without restrictions, as long as the data itself or the algorithm that generates it, does not leak sensitive information. 
}

\subsection{Related Work}

\emph{Generation of synthetic data.} Research in generative models has seen immense progress in the past decade. In particular, there has been a lot of interest in the  context of image processing for a myriad of tasks \cite{gui2020review}. In this context, GANs are one of the most popular frameworks, as they have been shown to produce more natural-looking images than the alternatives. \oldversion{Recurrent Neural Networks (RNNs)  \cite{sherstinsky_fundamentals_2018} and Long Short-Term Memory networks (LSTMs) \cite{hochreiter_long_1997} are most commonly used for treating sequential data.Apart from direct applications \cite{DBLP:journals/corr/abs-1803-06386,8317913,6637694}, they were also combined with GANs to generate sequential data \cite{doan_generating_2019,clark_efficient_2019,kundu_bihmp-gan:_2018}.} Previous works, \cite{chen2017modelfree,chen2018machine}, proposed GANs for the generation of power curves; however, issues of privacy were not addressed. Sequential models, such as RNNs \cite{sherstinsky_fundamentals_2018} and LSTMs \cite{hochreiter_long_1997}, are often used for forecasting; yet, our task is to generate new instances of data that imitate the properties of natural data, without being tied to a particular instance of the original samples. Therefore, we focus purely on GANs, as they implicitly capture the underlying distribution of natural data, and allow us to sample new instances.

\emph{Membership inference attacks.} ML algorithms can store features, or even entire records, that were used for its training \cite{fredrikson_model_2015,fredrikson_privacy_2014}. This means an attacker with access to an ML service could recover sensitive information about its training records. We want to prevent this and offer clean power consumption data for research or operational purposes. A minimum requirement for privacy is robustness to membership inference attacks. These attacks can identify members of the training set of an ML algorithm and use side information provided by the algorithm to determine properties of the identified client. Thorough analysis of these attacks is available in \cite{shokri_membership_2016,nasr_comprehensive_2018,truex_towards_2018}. However, little research has been done on the privacy leakage of generative models. A first example of a membership inference attack against a GAN was presented in \cite{hayes_logan:_2019}. Following that line of research, we propose several membership inference attacks to assess the privacy guarantees of our model. 

\subsection{Our contribution}

\textbf{GANs for sequential data generation:} The model proposed in this paper is specially adapted for sequential data. It utilizes 1-dimensional convolutional layers to capture the data structure.

\textbf{Evaluation using a forecasting task:} Assessing the quality of synthetic data is a highly non-trivial task. In the case of images, popular methods involve training well established classification models using synthetic data. These models are known to perform well when trained with natural data; thus, the idea is to use the performance drop when trained with synthetic data as a quality measure of the generated data \cite{borji_pros_2018}. In the case of power consumption curves, there are no well established classification models. Furthermore, in our setup, data is not naturally divided into easily distinguishable classes. Consequently, we contribute a quantitative method for assessing the quality of synthetic data by training an LSTM network for forecasting. The principle is the same as for images; we compare the performance of the same LSTM, when trained with natural and synthetic data.

\textbf{Evaluation using statistical properties of generated samples:} As the generated samples should accurately represent the households in the training set, we select statistical features of the power consumption curves, which are commonly utilized to infer information about the households \cite{beckel_towards_2012,hopf_feature_2014}. In particular, artificial curves should  closely resemble natural data in terms of their mean, coefficient of variation, skewness, kurtosis and maximum-mean ratio. The ensemble statistics of these ``{\indicators}'' on the generated data should be close to the actual statistics. As a quality metric, we use the earth mover's distance between the empirical distributions of the {\indicators} on artificial and natural curves.

\textbf{White-box Membership Inference Attacks:} In \cite{shokri_membership_2016, nasr_comprehensive_2018}, an attack using an inference model is proposed which  requires a dedicated training set. However, this may not be realistic and even unnecessary in some  scenarios. We propose simpler attacks that exploit similar structural properties of the network, but without the need of a dedicated training set for the attacker. Our first attack directly utilizes the output of the discriminator, as proposed in \cite{hayes_logan:_2019}, while the second attack utilizes the gradient of the loss function of the discriminator with respect to its model parameters, inspired by \cite{nasr_comprehensive_2018}. Both attacks compute a score for every input and select the top scoring input as the most likely candidate.

\textbf{Black-box Membership Inference Attacks:} All black-box attacks we have seen in the literature, including those proposed in \cite{shokri_membership_2016,nasr_comprehensive_2018,hayes_logan:_2019,truex_towards_2018}, are based on training a surrogate model that can be used to perform a white-box version of the attack. In general, this yields degraded results in comparison to the white-box attack. We propose to exploit the similarity between  artificial and natural data to perform a black-box attack. Our black-box attack is inspired by our evaluation method using statistical properties and uses the earth mover's distance between the empirical distributions of artificial and natural curves in terms of their {\indicators}.

\textbf{Defense Mechanisms:} We study the interplay between privacy and generalization, and how the choice of training hyper-parameters can protect against membership inference attacks. Additionally, we propose Gradient Norm Regularization, a defense mechanism designed to protect against membership attacks that exploit the gradient of the loss function.

\section{Privacy versus Utility Trade-offs}

\subsection{Membership Inference}

\oldversion{Privacy leakage is measured in terms of the accuracy of membership inference attacks.} \newversion{Membership inference consists of determining whether a particular example or set of examples were part of the training set of the target model.
  If an attacker cannot even determine membership, it is certainly not possible to obtain more detailed information. Thus, membership inference represents the weakest version of a violation of privacy.
  In our setup, we further simplify this task and provide 5 sets of examples to the attacker, asking for a prediction of which set was part of the training set.} The attacker knows that one and only one of these subsets was part of the training set. To determine the accuracy of the membership inference attacks, we repeat the attack and report the ratio of successful attempts. A more detailed explanation is given in Section~\ref{sec:experiments}. All the attacks presented in our work rely on computing a score for each subset, selecting the top scoring set as the most likely candidate. The rest of this section is dedicated to explaining the different scoring criteria.

\subsubsection{Likelihood Attack}

Recall that in the GAN setup, the job of the discriminator is to distinguish natural and synthetic samples. Intuitively, we predict that the discriminator will recognize samples, that were part of the training set, with higher confidence. The likelihood attack consists of querying the discriminator and using its average output (in $[0,1]$, where $0$: synthetic, $1$: natural) to assign a score to each set of curves.
\begin{algorithm}
\caption{Likelihood Attack}
\begin{algorithmic} 
\STATE \textbf{Input:} Universe $\mathcal{U}$, its partition $P_{\mathcal{U}}$, discriminator $D$.
\STATE \textbf{Output:} Index of the member of the training set $i$.
\STATE \textbf{initialize} $\mathrm{likelihoodList} \leftarrow []$
\FOR{\textbf{each} set \textbf{in} $P_{\mathcal{U}}$}
\STATE $\bar{p} \leftarrow$ \textbf{mean}$([D(\mathrm{curve})$ \textbf{for each} $\mathrm{curve}$ \textbf{in} set$])$
\STATE $\mathrm{likelihoodList}$\textbf{.append}$(\bar{p})$ 
\ENDFOR
\STATE \textbf{return} $\textbf{argmax}(\mathrm{likelihoodList})$
\end{algorithmic}
\label{algo:2}
\end{algorithm}
The likelihood attack is detailed in Algorithm \ref{algo:2}, where the average likelihood is denoted by $\bar{p}$. The attacker expects the discriminator to be more confident, i.e. output a higher likelihood, for curves that were seen during training. Thus, the attacker chooses the subset with the highest average likelihood. 

\subsubsection{Gradient Norm Attack}

Neural Networks (NNs) utilize iterative optimization algorithms, such as Stochastic Gradient Descent (SGD), to minimize a loss function that serves a training goal. SGD computes the gradient of the loss function with respect to the NN's parameters in order to minimize it. Since this is performed with training samples, the norm of the gradient of the loss function should be smaller at samples that were used during training, than on unseen samples. The gradient norm attack consists of computing the norm of the gradient of the loss function with respect to the model parameters and assigning a score to each set of natural curves according to their average gradient norm, then the attacker selects the subset with lowest average norm. This is a white-box attack, in which the attacker must have access to the internal parameters of the target model. Algorithm \ref{algo:1} details the attack.

\begin{algorithm}
\caption{Gradient-Norm Attack}
\begin{algorithmic} 
\STATE \textbf{Input:} Universe $\mathcal{U}$ with partition $P_{\mathcal{U}}$, generator $G$, discriminator $D_\theta$ with discriminator parameters $\theta$, Loss $L$.
\STATE \textbf{Output:} Index of the member of the training set $i$.
\STATE \textbf{initialize} $\mathrm{normList} \leftarrow []$ 
\FOR{\textbf{each} set \textbf{in} $P_{\mathcal{U}}$}
\STATE \textbf{initialize} $\mathrm{auxList} \leftarrow []$
\FOR{\textbf{each} $\mathrm{curve}$ \textbf{in} set}
\STATE $g \leftarrow \nabla_{\theta}L(D_\theta, G, \mathrm{curve})$
\STATE $\mathrm{auxList}$\textbf{.append}$(\|g\|_2)$
\ENDFOR
\STATE $\mathrm{normList}$\textbf{.append}$(\textbf{mean}(\mathrm{auxList}))$
\ENDFOR
\STATE \textbf{return} \textbf{argmin}($\mathrm{normList}$)
\end{algorithmic}
\label{algo:1}
\end{algorithm}

\subsubsection{Indicators Attack}

This black-box attack exploits the similarity between generated and natural data in terms of the {\indicators}. The idea behind this attack is that members of the training set will be closer to the synthetic samples in terms of these {\indicators}. For each subset of natural curves, the attacker fetches the same amount of artificial curves, computes the {\indicators} on both sets and compares them via the Earth Mover's Distance (EMD) of their empirical distributions. The set of natural curves with the smallest EMD is chosen. Let  $I$ be the vector-valued function that computes all {\indicators} for a set of curves. Algorithm \ref{algo:3} shows the algorithm for the indicators attack.

\begin{algorithm}
\caption{Indicators Attack}
\begin{algorithmic} 
\STATE \textbf{Input:} Universe $\mathcal{U}$, partition of universe ${P_{\mathcal{U}}}$, generator $G$, indicator functions $I$.
\STATE \textbf{Output:} Index of the member of the training set $i$.
\STATE \textbf{initialize} $\mathrm{distList} \leftarrow []$
\FOR{\textbf{each} set \textbf{in} ${P_{\mathcal{U}}}$}
\STATE $z \leftarrow$ \textbf{randN}($0,1$) \COMMENT{Random input for generator.}
\STATE $\mathrm{set}_{\mathrm{fake}} \leftarrow G(z)$
\STATE $\mathrm{AID} \leftarrow$ \textbf{mean}$($\textbf{EMD}$(I(\mathrm{set})$,$I(\mathrm{set}_\mathrm{fake})))$
\STATE $\mathrm{distList}$\textbf{.append}$(\mathrm{AID})$
\ENDFOR
\STATE \textbf{return} $\textbf{argmin}(\mathrm{distList})$
\end{algorithmic}
\label{algo:3}
\end{algorithm}

\subsection{Defense Mechanisms}
\label{subsec:def}

Membership inference attacks rely on overfitting. When this happens the target model memorizes and reproduces samples in its training set, instead of learning their properties and underlying distribution. The likelihood attack exploits bias in the discriminator output, while the gradient-norm attack exploits the local behavior of the loss function around the training samples. On the other hand, the indicators attack relies on the overly precise imitation of statistical properties of the training set, rather than proper generalization to the whole universe. This reasoning indicates that generalization could make the model more robust against the attacks presented here. To achieve good generalization, training hyper-parameters, such as the learning rates (LRs) for generator and discriminator, must be chosen carefully. Furthermore, the attacks could be incorporated in the training process and used to tune these hyper-parameters.

\subsubsection{Gradient Norm Regularization}

We propose a straight-forward way to combat the gradient-norm attack, by adding a term to the standard loss that penalizes a small norm of the gradient with respect to model parameters. More precisely, we suggest the regularized loss:
\begin{equation}
  L_{\mathrm{reg}}(\theta) = L_{\mathrm{discr}}(\theta) - \eta\Vert\nabla_{\theta}L_{\mathrm{discr}}(\theta)\Vert_2\;,
\end{equation}
where $L_{\mathrm{discr}}$ denotes the original discriminator loss, $\theta$ the parameters of the discriminator, and $\eta>0$ is a Lagrangian.

\section{Numerical Experiments}
\label{sec:experiments}

\newversion{
\subsection{Data Set}
}
Our experiments are based on ``SmartMeter Energy Consumption Data in London Households'' \cite{noauthor_smartmeter_nodate}. This set contains energy consumption readings for a sample of $5\,567$ London households that took part in the UK Power Networks led Low Carbon London project from November 2011 to February 2014. Power consumption readings were taken at half hour intervals in kWh (kilo Watt hours). For each reading a unique house identifier, date and time data and a CACI Acorn group label are provided. For our experiments we considered a mixed set of households with standard tariff from all CACI Acorn groups. We extracted two-week temporal frames, i.e., $672$ consecutive power consumption readings, which are separated according to each unique household identifier.

\newversion{
\subsection{Generative Model}
}

To generate synthetic power consumption curves we propose a 1-dimensional convolutional GAN. The generator network has a latent space of length 42. The latent space input is sampled from a normal distribution. This choice for the latent space dimension leaves enough room to capture the variation present in the training set. The generator network is composed of a combination of linear and 1D up-convolutional layers and uses ReLU activation functions, save for the last layer, which uses $\tanh$, to accord with the pre-processing of the data. The discriminator takes curves of length $672$ as inputs. It is composed of a combination of 1D convolutional and linear layers, followed by ReLU activation functions, save for the last one, where a sigmoid is used instead. To avoid the vanishing gradient problem and make training more stable, spectral normalization is applied to all the layers of the discriminator, as proposed in \cite{miyato_spectral_2018}. \newversion{The total complexity in terms of parameters is $1163909$ for the the generator and $1143769$ for the discriminator.}

\newversion{
\subsection{Evaluation Model}
\label{sec:lstm}
}

Synthetic data should provide the same utility as natural data when used for training an ML model for a certain task. We evaluate the quality of generated curves by training an LSTM model for forecasting using synthetic curves and testing its performance on a test set of natural curves. We train the same LSTM model using a portion of the training set of the generator and use this version of the model as a baseline for comparison. The LSTM architecture that we chose takes 24 points of the curve at a time as input and predicts the subsequent 24 points (12 hours). The output of the LSTM is post-processed with a linear layer and a $\tanh$ activation function. \newversion{The total complexity of this model is $15384$ parameters.} The performance of the LSTM is measured by Mean Squared Error (MSE) between prediction and ground truth on the test set. We compute the difference between the MSEs for the same LSTM, trained with artificial and natural curves. We refer to this difference as the LSTM score. The LSTM score can be negative when the LSTM trained with artificial curves performs better than the LSTM trained with natural curves.

For each curve we calculate the following indicators: mean, coefficient of variation, maximum-mean ratio, skewness, and kurtosis. The empirical distribution of each indicator over artificial and natural curves is computed. In our experiments we took the whole training set (around $2300$ examples) and the same amount of generated curves for this purpose. The Earth Mover's Distance (EMD) between the two histograms is used as a quality metric for evaluating the model. The average distance over all the different indicators is used to obtain a single quality index, which we call {\avgII}.  When computing the average distance over indicators, the indicators need to be comparable and hence require normalization. We normalize the indicators to unit variance, where the variance is estimated using both generated and natural data in order to prevent the normalization from distorting the result.

\subsection{Experimental procedure}

\begin{table*}
  \centering
  \begin{tabular}{ |c|c|c|c| } 
    \hline
    \textbf{Scenario} & i) \textbf{diff LR} & ii) \textbf{same LR}& iii) \textbf{Regularized (same LR)}  \\
    \hline
    Per-subset gradient-norm attack & $28.50\%$ & $68.50\%$ & $26.50\%$   \\ 
    Per-household gradient-norm attack & $21.77\%$ & $27.27\%$ & $20.68\%$ \\ 
    Per-subset likelihood attack & $39.00\%$ & $62.50\%$ & $39.50\%$ \\ 
    Per-household likelihood attack & $23.01\%$ & $27.33\%$ & $22.61\%$\\ 
    Indicators attack & $24.50\%$ & $33.00\%$ & $24.50\%$ \\
    \hline
    LSTM score & $0.011\pm{0.013}$ & $0.002\pm{0.007}$ & $0.016\pm{0.012}$\\
    Average Indicator Distance & $0.39\pm{0.14}$ & $0.29\pm{0.09}$ & $0.50\pm{0.18}$\\
    \hline
  \end{tabular}
  \caption{Accuracy of membership attacks, LSTM score and Average Indicator Distance for different setups.}
  \label{tab:attacks}
  \vspace{-1.5\baselineskip}
\end{table*}

To study the interplay between performance and privacy leakage, we train our generative model in three different scenarios: i) the generator has a significantly higher learning rate (LR); ii) generator and discriminator share the same LR; iii) gradient-norm regularization is applied during training and generator and discriminator share the same LR. The attacker, however, is allowed to exploit the full loss of the discriminator, including the regularization term. We observe the impact of the LR on the quality and privacy of the model.

For each run of our experiments, we completely shuffle the universe of natural curves. We obtain a partition by assigning each household to one of five subsets. Each subset contains the same number of households and a comparable amount of curves (around $2300$ each; the number of samples varies slightly from household to household). One subset is used for training and to assess the quality of the model in terms of the {\indicators}. The other four subsets will play the role of non-members when performing the attacks. One of these four subsets is also used as test set to evaluate the generator using the LSTM approach.

The generative model is trained using the GAN algorithm \cite{goodfellow_generative_2014}. Each run the model is trained for $140$ epochs with Adam \cite{kingma_adam:_2014}, $\mathrm{LR} = 10^{-4}$ for both generator and discriminator in the same LR (labeled ``same LR'') case and batch size $20$. In the case of different LR (labeled ``diff LR''), the LR of the generator is $10^{-4}$ and the LR of the discriminator is $10^{-5}$. For the regularized case, $\eta$ was set to $10^{-2}$.

To compute the LSTM score, we train two identical LSTMs: on training data and on the generated curves. Both LSTMs are trained for $40$ epochs using Adam \cite{kingma_adam:_2014}, with $\mathrm{LR} = 10^{-4}$ and batch size $50$. The loss function for training is a combination of the MSE and the $l_1$ norm of the difference between statistics of the prediction and the ground truth. The statistics considered for the loss are the mean and the second, third and fourth centralized moments of the curves. This combination is optimized in order to obtain the best performing LSTM. After training, both LSTMs are evaluated on the test set using the average Mean-Squared-Error. Then, the difference between the average MSEs of the fake LSTM (trained with artificial curves) and real LSTM (trained with natural curves) is the LSTM score. To compute the {\avgII}, we evaluate the {\indicators} (mean, coefficient of variation, maximum-mean ratio, skewness and kurtosis) for all curves in the training set, and for the same number of generated curves.
The Earth Mover's distance is evaluated between distributions of {\indicators} of natural and artificial curves, then normalized and averaged to obtain the {\avgII}.

The attacker is given 5 subsets and performs the attack on subsets as a whole, subsequently labeled ``per-subset''. The objective of the attacker is to determine which of the 5 subsets was the training set. In the ``per-household'' attack, a single household from each subset is selected randomly. The attacker is then given the data of these five households, one from each subset, and asked to determine which household was part of the training set. The per-subset attack are performed once per run, but per-household attacks are performed multiple times, by randomly picking different households.

\vspace*{.2\baselineskip}
\subsection{Results}

Table~\ref{tab:attacks} shows the success rate for different attacks  evaluated over ${\sim}100$ runs. Attacks were performed once per run in the per-subset case and $100$ times per run in the per-household case, randomly selecting different households each time. The {\indicators} attack was only performed on a per-subset basis. LSTM score and {\avgII} are reported for each subset with mean values over runs, and the error is the standard deviation. Having five subsets to choose from, a random guess would achieve $20\%$ accuracy. \newversion{Training the model with the same learning rate (ii)} achieves the best performance in terms of the LSTM score and {\avgII}; however, it also results in the most vulnerable model in terms of privacy. \newversion{Modifying the LR (i)} results in significant gains in privacy, at the cost of marginal loses in performance.

\begin{figure}
  \centering
  \begin{tikzpicture}
    \begin{axis}[poweraxisHalf]
      \addplot +[powerplot] table [powertable0]{plots/Forecast9.csv};
      \addlegendentry{Real}
      \addplot +[powerplot] table [powertable1]{plots/Forecast9.csv};
      \addlegendentry{Predicted}
    \end{axis}
  \end{tikzpicture}
  \caption{Prediction by a model trained with artificial curves.}
  \label{fig:1}
\end{figure}
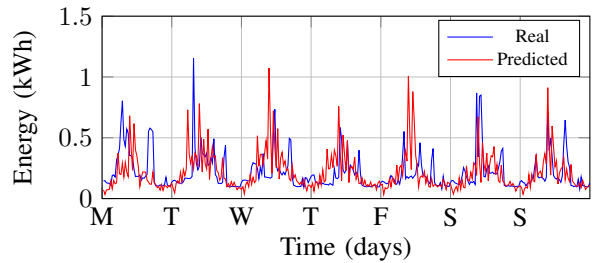

As expected, per-household attacks are overall weaker than per-subset attacks. This means that these attacks depend on the ensemble statistics of large groups of samples in order to be effective. \newversion{Gradient-norm regularization (iii)} provides additional robustness against the gradient-norm attack, but the sacrifice in performance is high compared to the slim gain that could be obtained by simply tuning the LR, as in (i). Likelihood attacks are more robust against defense mechanisms and provide a $20\%$ gain over a random guess, even in unfavorable conditions for the attacker. On the other hand, the gradient-norm attack can perform better in poor training conditions, but is easier to mitigate.

\begin{figure}
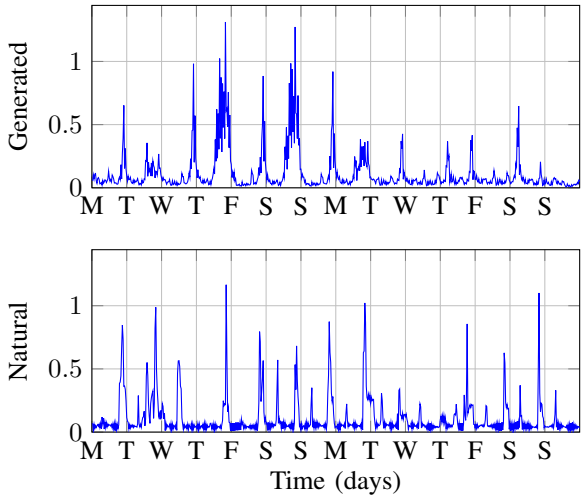

  \centering
  \powerplotComparison{2}    
  \caption{Two generated weeks of power consumption data (in kWh), as well as the closest (in terms of Earth Mover's Distance) curve in the training set.}
  \label{fig:2}
\end{figure}

Fig. \ref{fig:1} shows predictions performed by an LSTM network on a curve from the test set. The LSTM model always predicts half a day (24 points) from the preceding week (336 point) of ground truth data at a time. The plot results from joining these predictions together. Fig.  \ref{fig:2} showcases a random artificial power curve generated by the same LR generative model. Below it is the closest natural curve from the model's training set, in terms of their EMD.

\section{Summary and Concluding Remarks}

We provided a privacy-aware method for power consumption data generation. Additionally, we proposed two novel methods for evaluation of GANs for sequential data. Firstly, the LSTM score, which uses the forecasting performance of an LSTM and can be generalized to all types of sequential data. Secondly, the {\avgII} which uses the statistics of power consumption curves. To this end, we chose features that are useful for classification tasks on power consumption data \cite{beckel_towards_2012,hopf_feature_2014}.

We studied the interplay between privacy and quality in generative networks for power consumption data. We found that there is a trade-off between the quality of the generated samples and the privacy guarantees of the generative model. When using the discriminator for attacks, the success rates are similar to those found for classifiers \cite{shokri_membership_2016,nasr_comprehensive_2018,truex_towards_2018}. Yet, with proper training setup, the model can be robust against membership inference attacks, which provides a strong guarantee against other attacks, like attribute inference.

Our model can be improved and adapted to other types of power consumption data. For example, by taking readings over different time spans or conditioning on weather/temperature. Our method for data generation and privacy analysis allows power companies to share power consumption data in an useful and privacy-preserving way.

\section*{Acknowledgment}
This research was supported by the Gaspard Monge program for Optimization and Operations Research (PGMO). This project has received funding from the European Union’s Horizon 2020 research and innovation programme under the Marie Skłodowska-Curie grant agreement No 792464.

{\small
\bibliographystyle{ieee_fullname}
\bibliography{main.bib}
}

\end{document}